# Enhanced beam-beam modeling to include longitudinal variation during weak-strong simulation

Derong Xu[1,*], Vasiliy S. Morozov[2], David Sagan[3], Yue Hao[4], and Yun Luo[1]

[1]*Brookhaven National Laboratory, Upton, New York, USA*
[2]*Oak Ridge National Laboratory, Oak Ridge, Tennessee, USA*
[3]*Cornell University, Ithaca, New York, USA*
[4]*Michigan State University, East Lansing, Michigan, USA*

Beam-beam interactions pose substantial challenges in the design and operation of circular colliders, significantly affecting their performance. In particular, the weak-strong simulation approach is pivotal for investigating single-particle dynamics during the collider design phase. This paper evaluates the limitations of existing models in weak-strong simulations, noting that while they accurately account for energy changes due to slingshot effects, they fail to incorporate longitudinal coordinate changes ($z$ variation). To address this gap, we introduce two novel transformations that enhance Hirata's original framework by including both $z$ variation and slingshot effect-induced energy changes. Through rigorous mathematical analysis and extensive weak-strong simulation studies, we validate the efficacy of these enhancements in achieving a more precise simulation of beam-beam interactions. Our results reveal that although $z$ variation constitutes a higher-order effect and does not substantially affect the emittance growth rate within the specific design parameters of the Electron-Ion Collider, the refined model offers improved accuracy, particularly in scenarios involving the interaction between beam-beam effects and other random diffusion processes, as well as in simulations incorporating realistic lattice models.

## I. INTRODUCTION

Luminosity, the primary metric in accelerator physics for a collider, measures the rate of physics events per unit cross section per second during beam collisions. Achieving higher peak and integrated luminosity is pursued in colliders across energy [1], precision [2], and quantum chromodynamics frontiers [3].

In circular colliders, beam-beam interactions—the electromagnetic forces between opposing beams—significantly limit performance. For lepton colliders, this interaction introduces a "beam-beam limit," where luminosity increases below the expected relationship with beam intensity beyond a certain threshold [4]. In hadron colliders, beam emittance growth, primarily driven by these interactions, curtails luminosity lifetime, as demonstrated in empirical studies [5,6]. The "beam-beam parameter," quantifying the maximum tune shift during collisions, serves as a key metric for assessing interaction strength.

The Electron-Ion Collider (EIC), to be constructed at Brookhaven National Laboratory, aims to achieve an unprecedented luminosity of $10^{34}$ cm$^{-2}$ s$^{-1}$ by colliding 10 GeV electrons and 275 GeV protons, requiring large electron and proton beam-beam parameters [7]. Throughout the design phase of the EIC, a critical objective is to minimize proton emittance growth amidst beam-beam interaction.

Hadron–Elektron Ring Anlage (HERA) pioneered the collision of leptons and protons, serving as the first operational collider of its kind. However, in comparison to the main parameters achieved during routine operation at HERA [8], the EIC seeks to escalate the peak luminosity objective by two orders of magnitude, accompanied by a fourfold increase in both proton and electron beam-beam parameters. Given the unprecedented nature of the EIC's design objective, beam-beam simulations emerge as the sole method to validate the efficacy of the parameter combination, with accurate modeling of beam-beam interaction being crucial for assessing proton emittance growth in the simulation.

In beam-beam simulations, two principal approaches are distinguished: strong-strong and weak-strong. The strong-strong concept was introduced by Peggs in the 1980s [9]. In this paper, the strong beam distribution is fixed during tracking, and the role of the strong beam were switched between the interacting beams over successive tracking turns. Modern strong-strong simulations have advanced to a point where the beam distributions are dynamically updated based on the electromagnetic fields calculated from a two-dimensional Poisson equation. Subsequently, these distributions evolve according to the Vlasov equation, ensuring a self-consistent simulation process.

In contrast to the strong-strong approach, weak-strong simulations simplify the modeling of beam-beam interactions by assuming the strong beam maintains a rigid Gaussian distribution. This simplification facilitates the calculation of electromagnetic forces exerted on test particles within the weak beam using the Bassetti-Erskine formula [10], a method that, despite lacking self-consistency and the ability to address coherent effects, remains crucial for collider design due to its lower computational demands. Particularly in hadron storage ring design, where effective cooling methods are absent and particles must be tracked over millions of turns to ascertain stability, the weak-strong model proves indispensable. Moreover, it introduces less numerical noise compared to the strong-strong method, which employs the particle-in-cell technique to expedite Poisson equation solutions [11–13]. This technique, while efficient, often leads to significant numerical noise by projecting particle distributions onto a finite two-dimensional grid, as observed during the EIC design phase [14–18]. Such noise can obscure the particle diffusion effects attributed to beam-beam interactions, making them discernible only through weak-strong simulations.

For the EIC, the adoption of a flat hadron beam configuration—characterized by a vertical emittance that is an order of magnitude smaller than horizontal emittance—maximizes luminosity while introducing sensitivity to various real-world fluctuations in the vertical plane. The weak-strong simulation has emerged as a cornerstone in addressing EIC design challenges, including the dynamic aperture reduction from interaction region (IR) magnetic field errors [19], emittance growth due to electron orbit perturbations from dipole magnet power supply variability [20], and crab cavity phase noise [21,22]. The necessity for a precise weak-strong model transcends the EIC, underscoring its critical importance across the collider physics community for ensuring the integrity of collider design.

This paper is organized as follows. Section II revisits the concept of synchro-beam mapping, highlighting the $z$ variation effect, and the energy changes resulting from slingshot effects. Section III introduces two symplectic approaches designed to encompass both the $z$ variation effect and the slingshot effect-induced energy changes. Section IV provides a comparative analysis of the simulation outcomes using these three distinct methodologies. A conclusive summary is presented in Sec. V.

## II. SYNCHRO-BEAM MAPPING

A single-particle dynamics is described by the set of canonical coordinates:

$$\mathbf{x} = (x, p_x, y, p_y, z, p_z), \quad (1)$$

where $x$ and $y$ represent the transverse positions in the horizontal and vertical planes, respectively, and $p_{x,y}$ are their associated momenta, normalized to the design momentum, $P_0$, of the reference particle. The longitudinal position is given by $z = s - l$, where $s$ is the designed path length for the reference particle, and $l$ is the actual path length. The term $p_z = (P - P_0)/P_0$ denotes the deviation in momentum relative to the reference particle. For the purposes of clarity and focus, this paper restricts its scope to the case of high-relativistic dynamics.

The choice of canonical coordinates and the Hamiltonian is discussed in Appendix A.

### A. Hirata's original approach

The synchro-beam mapping, formulated by Hirata, Moshammer, and Ruggiero, is extensively utilized for simulating beam-beam interactions in the presence of synchrotron motion, providing a symplectic mapping within a six-dimensional phase space [23]. This method involves longitudinally splitting the strong bunch into multiple slices and employing a drift-kick-drift model to calculate the particle-slice interaction. In this approach, the particle engages in a virtual drift from the interaction point (IP) to the collision point (CP), receives a beam-beam kick from the opposing slice at the CP, and then returns back to the IP.

Denoting $z^*$ the longitudinal coordinate of the strong slice in its own axis, the separation between the IP and CP is represented as

$$S(z, z^*) = \frac{z - z^*}{2}. \quad (2)$$

Throughout this paper, we will use $S$ to represent $S(z, z^*)$ without causing any ambiguity.

The virtual drift from the IP to the CP is characterized by an exponential Lie operator $\mathcal{D}_0$, defined as

$$\mathcal{D}_0 = \exp\left(-:SH_0:\right), \quad \text{where } H_0 = \frac{p_x^2 + p_y^2}{2}. \quad (3)$$

This operator's action on the canonical coordinates results in the transformation:

$$\begin{aligned}
\mathcal{D}_0 x &= x + p_x S, & \mathcal{D}_0 p_x &= p_x, \\
\mathcal{D}_0 y &= y + p_y S, & \mathcal{D}_0 p_y &= p_y, \\
\mathcal{D}_0 z &= z, & \mathcal{D}_0 p_z &= p_z - \frac{p_x^2 + p_y^2}{4}.
\end{aligned} \quad (4)$$

The beam-beam interaction at the CP is represented by another Lie operator $\mathcal{B}$, expressed as

$$\mathcal{B} = \exp(-:U:), \quad (5)$$

where $U$ denotes the beam-beam potential induced by the strong slice. The application of the operator $\mathcal{B}$ to the coordinates results exclusively in modifications to the momenta, described by

$$(\Delta p_x, \Delta p_y, \Delta p_z) = -\nabla U. \quad (6)$$

The spatial coordinates remain unaffected by the beam-beam interaction, given the assumption that the slice is sufficiently "thin." Transverse momentum changes, $\Delta p_{x,y}$, are derived from the well-established Bassetti-Erskine formula, and the formulation for the longitudinal momentum change, $\Delta p_z$, follows Hirata's proposal in [23]. Appendix B provides detailed formulas for momentum changes by a strong slice characterized by a bi-Gaussian distribution.

Upon experiencing the beam-beam kick at the CP, the test particle is subjected to another virtual drift, leading it back from the CP to the IP. This sequential process is encapsulated by the overall mapping equation:

$$\mathcal{M}_0 = \mathcal{D}_0 \mathcal{B} \mathcal{D}_0^{-1}, \quad (7)$$

which represents the combined effect of the initial drift, the beam-beam interaction, and the subsequent return drift. The resultant transformation of the particle's coordinates and momenta is given by

$$x^{\text{new}} = x - S\Delta p_x,$$
$$p_x^{\text{new}} = p_x + \Delta p_x,$$
$$y^{\text{new}} = y - S\Delta p_y,$$
$$p_y^{\text{new}} = p_y + \Delta p_y,$$
$$z^{\text{new}} = z,$$
$$p_z^{\text{new}} = p_z + \Delta p_z - \frac{p_x^2 + p_y^2}{4} + \frac{(p_x^{\text{new}})^2 + (p_y^{\text{new}})^2}{4}, \quad (8)$$

where the new positions and momenta are adjusted according to the shifts induced by the beam-beam interaction.

### B. z variation and slingshot effects

Hirata's approach provides a symplectic mapping for the beam-beam interaction, which includes the energy change and the bunch-length effect. This approach employs exponential Lie operators to represent both the virtual drift and the beam-beam kick, ensuring that each component of the synchro-beam mapping maintains symplecticity. Consequently, the cumulative mapping that results from sequentially applying these steps preserves the symplecticity.

However, it is imperative to acknowledge a critical distinction: the transformation characterizing the transition between the IP and the CP does not retain symplecticity when considered in the context of the standard accelerator coordinates $(x, p_x, y, p_y, z, p_z)$. This discrepancy arises from the dynamic nature of the CP location, which varies among particles based on their initial coordinates at the IP. Specifically, some particles engage in collisions at CP1, others at CP2, and so forth, necessitating the computation of particle coordinates at their respective CPs. Once all particles have transitioned to their respective CPs, the overarching transformation from the IP to these dynamically defined CPs does not conform to symplectic principles.

To illustrate this concept, we can incorporate Eq. (2) into an actual drift map. Consider the simple drift Hamiltonian given by $H_0 = (p_x^2 + p_y^2)/2$. The transformation corresponding to a real drift over a distance $S$ is described by

$$x = x_0 + p_{x,0}S, \qquad p_x = p_{x,0},$$
$$y = x_0 + p_{y,0}S, \qquad p_y = p_{y,0},$$
$$z = z_0, \qquad p_z = p_{z,0}.$$

The corresponding Jacobian matrix is given by

$$M = \frac{\partial(x, p_x, y, p_y, z, p_z)}{\partial(x_0, p_{x,0}, y_0, p_{y,0}, z_0, p_{z,0})}$$

$$= \begin{bmatrix} 1 & S & 0 & 0 & p_{x,0}/2 & 0 \\ 0 & 1 & 0 & 0 & 0 & 0 \\ 0 & 0 & 1 & S & p_{y,0}/2 & 0 \\ 0 & 0 & 0 & 1 & 0 & 0 \\ 0 & 0 & 0 & 0 & 1 & 0 \\ 0 & 0 & 0 & 0 & 0 & 1 \end{bmatrix}.$$

It is straightforward to verify that

$$M^T J M = \begin{bmatrix} 0 & 1 & 0 & 0 & 0 & 0 \\ -1 & 0 & 0 & 0 & -p_{x,0}/2 & 0 \\ 0 & 0 & 0 & 1 & 0 & 0 \\ 0 & 0 & -1 & 0 & -p_{y,0}/2 & 0 \\ 0 & p_{x,0}/2 & 0 & p_{y,0}/2 & 0 & 1 \\ 0 & 0 & 0 & 0 & -1 & 0 \end{bmatrix} \neq J,$$

where $J$ is the 6-by-6 symplectic form matrix.

This deviation from symplecticity can be intuitively understood through an analogy to photography in phase space. According to Liouville's theorem, in a closed and isolated system, where dynamics are governed by Hamiltonian principles without any dissipation or diffusion, the volume occupied by the system in phase space remains invariant over time. When all particles are captured in a single "photograph," the phase volume should remain constant. However, when we capture only a subset of particles in sequential "photographs," the volume of phase space represented in each image may vary depending on the distribution evolution. Consequently, when summing the volumes from these sequential photographs taken at different times, the total phase volume may not remain constant.

Easing off the symplectic constraint, Fig. 1 presents the actual drift of a particle between the IP and the CP. The respective path lengths are quantified as follows:

$$l_{\text{IP}\to\text{CP}} = \frac{S}{\sqrt{1 - x'^2 - y'^2}}, \quad (9)$$

$$l_{\text{CP}\to\text{IP}} = \frac{S}{\sqrt{1 - (x' + \Delta x')^2 - (y' + \Delta y')^2}}, \quad (10)$$

where $x' = dx/ds$ and $y' = dy/ds$ denote the derivatives of $x$ and $y$ with respect to the path length $s$, while $\Delta x'$ and $\Delta y'$ represent the deflections due to the beam-beam interaction at the CP. This difference in path length accounts for the alteration in the $z$ coordinate as the test particle returns to the IP:

$$\Delta z = l_{\text{CP}\to\text{IP}} - l_{\text{IP}\to\text{CP}}$$
$$\approx \frac{S}{2}[(x' + \Delta x')^2 + (y' + \Delta y')^2 - x'^2 - y'^2]. \quad (11)$$

In this paper, the longitudinal coordinate change is referred to as $z$ variation effect.

When every particle undergoes a beam-beam kick at the CP and returns to the IP, the entire process as a whole must retain symplecticity. However, given the actual drift between the IP and the CP deviates from symplectic behavior, the beam-beam interaction at the CP must also be nonsymplectic in nature to ensure the overall process conserves symplecticity.

In addition to the momentum change or energy change induced by distribution variation, as shown in Eq. (6), there is an additional mechanism of energy alteration reminiscent of the gravitational slingshot effect encountered in orbital mechanics, as explained in [24]. The comprehensive derivation of this effect is provided in Appendix C. In terms of the accelerator coordinates, the momentum change due to the slingshot effect is

$$(\Delta p_z)_{\text{slingshot}} = \frac{(x' + \Delta x')^2 + (y' + \Delta y')^2 - x'^2 - y'^2}{4}. \quad (12)$$

By comparing Eqs. (11) and (12) with Eq. (8), we observe that while Hirata's approach encapsulates the energy change attributable to the slingshot effect, it does not account for $z$ variation. In the original formulation by Hirata, the $z$ coordinate is presumed to remain unchanged throughout the interaction. This assumption, however, does not fully capture the dynamics of the system. The effect of $z$ variation may have significant implications in the design, especially given that the hadron particles are often tracked over millions of turns. Recognizing and addressing this $z$ variation is crucial for accurately modeling the long-term behavior of particles within accelerators, pointing to areas where existing theoretical frameworks might be enhanced to better reflect physical realities.

## III. MODIFIED VIRTUAL DRIFT

Despite the nonsymplectic nature of the actual transformation from the IP to the CP in practical scenarios, Hirata's approach retains its significance for two key reasons: first, it ensures the symplecticity of the overall synchro-beam mapping, and second, it accurately captures the slingshot effects. Building on this foundation, we aim to extend Hirata's methodology in this section by devising an approximate yet symplectic transformation for the virtual drift. This transformation will account for both $z$ variation and slingshot effects at the lowest order, thereby refining the model to more closely mirror the complexities observed in Sec. II B.

### A. Chromatic Hamiltonian

Incorporating $p_z$ into the Hamiltonian expansion allows for calculation of $z$ variation effects. This leads to the refined Hamiltonian:

$$H_1 = \frac{p_x^2 + p_y^2}{2(1 + p_z)}. \quad (13)$$

Following Hirata's strategy, the associated exponential Lie operator is defined as

$$\mathcal{D}_1 = \exp(:h_1:) \quad \text{where } h_1 = -SH_1. \quad (14)$$

We remind the reader that the Lie operator is a shorthand notation for the Poisson bracket. For any two functions of $f$ and $g$, the Poisson bracket is defined as

$$:f:g = [f, g] = \sum_{u=x,y,z} \frac{\partial f}{\partial u}\frac{\partial g}{\partial p_u} - \frac{\partial f}{\partial p_u}\frac{\partial g}{\partial u}. \quad (15)$$

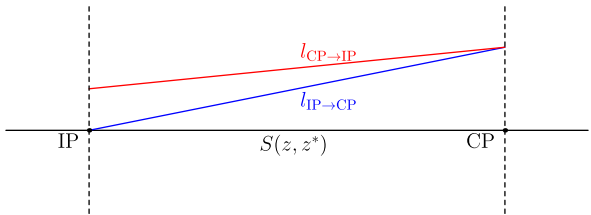

FIG. 1. Illustration of test particle drift between the IP and CP. The trajectory from IP to CP is depicted by the blue line, whereas the return path from CP to IP is represented by the red line.

Application of the $:h_1:$ operator to longitudinal coordinates

$$:h_1:z = [-SH_1, z] = S\frac{\partial H_1}{\partial p_z} = \frac{h_0}{(1+p_z)^2},$$
$$:h_1:p_z = [-SH_1, p_z] = -H_1\frac{\partial S}{\partial z} = \frac{h'_0}{1+p_z}, \quad (16)$$

where

$$h_0 = -H_0 S = -\left(\frac{p_x^2 + p_y^2}{2}\right)S,$$
$$h'_0 = -H_0\frac{\partial S}{\partial z} = -\left(\frac{p_x^2 + p_y^2}{2}\right)S'. \quad (17)$$

Iterative applications of the operator reveal

$$:h_1:^n z = \frac{S}{S'}\left(-\frac{1}{2}\right)_n \frac{(2S'H_0)^n}{(1+p_z)^{2n}},$$
$$:h_1:^n p_z = \left(-\frac{1}{2}\right)_n \frac{(2S'H_0)^n}{(1+p_z)^{2n-1}}, \quad (18)$$

for $n \geq 1$. Here $(\cdot)_n$ is the Pochhammer symbol of rising factorial

$$(q)_n = q(q+1)(q+2)\cdots(q+n-1). \quad (19)$$

The resulting transformations for $z$ and $p_z$ are

$$\mathcal{D}_1 z = z + \sum_{n=1}^{\infty} \frac{:h_1:^n z}{n!}$$
$$= z - \frac{S}{S'} + \frac{S}{S'}{}_1F_0\left(-\frac{1}{2};;\frac{2S'_0 H_0}{(1+p_z)^2}\right),$$
$$\mathcal{D}_1 p_z = p_z + \sum_{n=1}^{\infty} \frac{:h_1:^n p_z}{n!}$$
$$= -1 + (1+p_z){}_1F_0\left(-\frac{1}{2};;\frac{2S'H_0}{(1+p_z)^2}\right), \quad (20)$$

where ${}_1F_0$ is the generalized hyper-geometric function ${}_mF_n$ with $m=1$, $n=0$.

Utilizing the identity

$${}_1F_0(a;;x) = \frac{1}{(1-x)^a}, \quad (21)$$

the transformation of the virtual drift from the IP to the CP can be expressed as

$$\Phi(p_x, p_y, p_z) = \sqrt{1 - S' \cdot \frac{p_x^2 + p_y^2}{(1+p_z)^2}} - 1,$$
$$\mathcal{D}_1 x = x + \frac{Sp_x}{1+p_z}, \quad \mathcal{D}_1 p_x = p_x,$$
$$\mathcal{D}_1 y = y + \frac{Sp_y}{1+p_z}, \quad \mathcal{D}_1 p_y = p_y,$$
$$\mathcal{D}_1 z = z + \frac{S}{S'}\Phi(p_x, p_y, p_z),$$
$$\mathcal{D}_1 p_z = p_z + (1+p_z)\Phi(p_x, p_y, p_z). \quad (22)$$

The transformation from the CP to the IP reverses the effects of $\mathcal{D}_1$, achieved by inverting the sign of $S$ and $S'$:

$$\Psi(p_x, p_y, p_z) = \sqrt{1 + S' \cdot \frac{p_x^2 + p_y^2}{(1+p_z)^2}} - 1,$$
$$\mathcal{D}_1^{-1} x = x - \frac{Sp_x}{1+p_z}, \quad \mathcal{D}_1^{-1} p_x = p_x,$$
$$\mathcal{D}_1^{-1} y = y - \frac{Sp_y}{1+p_z}, \quad \mathcal{D}_1^{-1} p_y = p_y,$$
$$\mathcal{D}_1^{-1} z = z + \frac{S}{S'}\Psi(p_x, p_y, p_z),$$
$$\mathcal{D}_1^{-1} p_z = p_z + (1+p_z)\Psi(p_x, p_y, p_z). \quad (23)$$

The operator $\mathcal{D}_1$ in Eq. (14) implies that the location of the CP is determined by the longitudinal coordinate $z$ at the IP. During the drift from the IP to the CP, $z$ is no longer a constant, as shown in Eq. (22). This variation means that the test particle may not engage with the counter-propagating slice at the initially calculated CP location. To effectively apply the collision map $\mathcal{B}$, delineated in Eq. (5), we have to presume that the $z$ variation—arising from the modified virtual drift—exerts negligible influence on the collision map $\mathcal{B}$.

This assumption, despite not being entirely realistic, is not anticipated to introduce significant discrepancies. The electromagnetic field generated by a relativistic beam is characterized by a narrow angular width of $1/\gamma^*$, where $\gamma^*$ denotes the Lorentz factor of the strong slice. Let $\sigma$ represent the transverse size encountered by a test particle interacting with this strong slice. The test particle will predominantly experience the electromagnetic field from the thin slice if:

$$\Delta z \ll \sigma/\gamma^*. \quad (24)$$

In most practical scenarios, this condition is generally met.

### B. Exact Hamiltonian

By applying an exponential Lie operator to obtain the virtual drift transformation, we have extended Hirata's methodology for the chromatic drift Hamiltonian.

However, it proves to be impractical for the exact drift Hamiltonian due to the resulting expression's complexity and unwieldiness. Therefore, it is imperative to explore an alternative strategy to effectively tackle this challenge.

In Sec. III A, it is demonstrated that the CP location is not solely determined by the initial longitudinal coordinates. Here, we continue to use $S$ to represent the drift length from the IP to the CP. To account for geometric effects, $S$ must be contingent on the transverse momenta $p_{x,0}$ and $p_{y,0}$, and to consider the chromatic effect, $S$ is also influenced by $p_{z,0}$. The superscript "0" means these quantities are evaluated at the IP. As a result, the relationship governing $S$ can be expressed as follows:

$$S = S(z_0, z^*, p_{x,0}, p_{y,0}, p_{z,0}). \quad (25)$$

The variable $z^*$ remains constant when the thin slice traverses between the IP and the CP.

At the CP, the test particle has a longitudinal coordinate $z$. Although the expression of $z$ is still unknown yet, $S$ can be expressed by $z$ and $z^*$. The elapsed time for the reference particle from the IP to the CP is $t_r = S/v_r$. The superscript "$r$" indicates that $v_r$ is the velocity of the reference particle. Using the definition of $z$ in Eq. (A7),

$$z = s_r - l = v_r t_r - vt \Rightarrow t = \frac{v_r t_r - z}{v} = \frac{S - z}{v}, \quad (26)$$

where $t$ is when the test particle arrives at CP. For the opposite slice, the arrival time is obtained by substituting $S$ with $-S$, where the negative sign comes from the opposite $s$ axis:

$$t^* = \frac{-S - z^*}{v^*}. \quad (27)$$

The test particle will collide with the slice when $t = t^*$:

$$\frac{S - z}{v} = \frac{-S - z^*}{v^*} \Rightarrow S(z, z^*) = \frac{v^* z - v z^*}{v + v^*}. \quad (28)$$

Taking the relativistic limit $v = v^* = c$, Eq. (28) will turn into Eq. (2). It is crucial to note, however, that the interpretation of the variable $z$ varies between the two equations. In Eq. (28), $z$ represents the longitudinal coordinate at the CP, whereas in Eq. (2), the same symbol $z$ actually denotes the quantity $z_0$ determined at the IP.

As a result, $S$ has a form of

$$\begin{aligned} S &= S(z_0, z^*, p_{x,0}, p_{y,0}, p_{z,0}) \\ &= S(z(z_0, p_{x,0}, p_{y,0}, p_{z,0}), z^*). \end{aligned} \quad (29)$$

The spatial coordinate transformation governed by the exact drift Hamiltonian is detailed as

$$\begin{aligned} x &= x_0 + \left(\frac{p_{x,0}}{p_{s,0}}\right) S(z, z^*), \\ y &= y_0 + \left(\frac{p_{y,0}}{p_{s,0}}\right) S(z, z^*), \\ z &= z_0 - \left(\frac{H_0}{p_{s,0}}\right) S(z, z^*), \end{aligned} \quad (30)$$

where

$$p_{s,0} = \sqrt{(1 + p_{z,0})^2 - p_{x,0}^2 - p_{y,0}^2} = 1 + p_{z,0} - H_0. \quad (31)$$

The variable $z$ on the right-hand side in Eq. (30) is evaluated at the CP, and $z$ can be resolved by combining Eqs. (28) and (30).

We introduce the generating function to find out the transformation of $p_{x,y,z}$:

$$G_3 = -x p_{x,0} - y p_{y,0} - z p_{z,0} + H_0 S(z, z^*). \quad (32)$$

The spatial coordinates at the IP are given by

$$x_0 = -\frac{\partial G_3}{\partial p_{x,0}}, \quad y_0 = -\frac{\partial G_3}{\partial p_{y,0}}, \quad z_0 = -\frac{\partial G_3}{\partial p_{z,0}}. \quad (33)$$

This relationship confirms the alignment of the spatial coordinate transformation equations with those previously established in Eq. (30).

The transformation of momentum coordinates at the CP is similarly derived from the partial derivatives of $G_3$ with respect to the spatial coordinates, yielding:

$$\begin{aligned} p_x &= -\frac{\partial G_3}{\partial x} = p_{x,0}, \\ p_y &= -\frac{\partial G_3}{\partial y} = p_{y,0}, \\ p_z &= -\frac{\partial G_3}{\partial z} = p_{z,0} - H_0 \frac{\partial S(z, z^*)}{\partial z}. \end{aligned} \quad (34)$$

Similar to the Hirata's approach, the transverse momenta remain unchanged, and the correction is applied to $p_z$ to preserve the symplectic structure of the system.

The reverse transformation, from the CP back to the IP, is achieved by representing the initial coordinate and momentum set $(x_0, p_{x,0}, y_0, p_{y,0}, z_0, p_{z,0})$ in terms of the final set $(x, p_x, y, p_y, z, p_z)$, utilizing a combination of Eqs. (30) and (34).

For relativistic case, the transformation from the IP to the CP is summarized as

$$S = \frac{(z_0 - z^*)p_{s,0}}{2p_{s,0} + H_0},$$

$$x = x_0 + \left(\frac{p_{x,0}}{p_{s,0}}\right)S, \qquad p_x = p_{x,0},$$

$$y = y_0 + \left(\frac{p_{y,0}}{p_{s,0}}\right)S, \qquad p_y = p_{y,0},$$

$$z = z^* + 2S, \qquad p_z = p_{z,0} - \frac{H_0}{2}, \qquad (35)$$

and the reverse transformation from the CP to the IP:

$$H_0 = \frac{p_x^2 + p_y^2}{2(1 + p_z)}, \quad p_{s,0} = 1 + p_z - \frac{1}{2}H_0, \quad S = \frac{z - z^*}{2},$$

$$p_{x,0} = p_x, \qquad x_0 = x - \left(\frac{p_x}{p_{s,0}}\right)S,$$

$$p_{y,0} = p_y, \qquad y_0 = y - \left(\frac{p_y}{p_{s,0}}\right)S,$$

$$p_{z,0} = p_z + \frac{1}{2}H_0, \qquad z_0 = z + \left(\frac{H_0}{p_{s,0}}\right)S. \qquad (36)$$

## IV. SIMULATION

Section III details the derivation of the enhanced virtual drift, incorporating $z$ variation and energy adjustments due to slingshot effects. Weak-strong simulations are conducted to validate the enhanced map's accuracy and assess its implications, ensuring the theoretical modifications are both substantiated and practically relevant.

### A. Model difference demonstration

To demonstrate the distinctions among the three models, simulations are performed using artificially constructed parameters. These "man-made" parameters, which do not correspond to real-world machines, are specifically designed to highlight the differences between the models. The beam-beam parameters are set at 0.01, with the associated beam characteristics presented in Table I.

Figure 2 illustrates the variation in the longitudinal coordinate $z$ after the test particle undergoes a virtual drift from the IP to the CP, receives a beam-beam kick at CP, and drifts back to IP. The initial coordinate of the test particle is set at $(\sigma, 0, \sigma, 0, 0, 0)^T$, where $\sigma = 70$ μm. As the location of the thin slice changes, so does $\Delta z$. For Hirata's model, $\Delta z = 0$. In both the chromatic and exact models, $\Delta z$ is aligned with each other but remains small with a magnitude of $10^{-11}$ m.

The $z$ variation is too small to impact beam dynamics observably when the beam-beam kick is not sufficiently large. Figure 3 depicts the frequency diffusion over the course of tracking 1000 turns. The thin slice is positioned at $z^* = -\beta/2 = 30$ cm. The initial particle distribution spans $\pm 5\sigma$ across the $x$–$y$ plane, with the longitudinal coordinates and $p_{x,y}$ set to zero. The diffusion index $D$ is determined using the formula:

TABLE I. Artificial parameters used to illustrate $z$ variations among different virtual drift models. The transverse working point is based on the SuperKEKB design [25].

| Parameter | Unit | Proton and antiproton |
|---|---|---|
| Energy | GeV | 275 |
| Particles per bunch | $10^{11}$ | 2.1 |
| $\beta_x$ at IP | cm | 60.0 |
| $\beta_y$ at IP | cm | 60.0 |
| rms bunch size (H/V) | μm | 70.0 |
| Horizontal fractional tune | ⋯ | 0.530 |
| Vertical fractional tune | ⋯ | 0.570 |
| Synchrotron tune | ⋯ | −0.001 |

$$D = \log_{10}\sqrt{(\nu_{x,2} - \nu_{x,1})^2 + (\nu_{y,2} - \nu_{y,1})^2}. \qquad (37)$$

Here, $\nu_1$ and $\nu_2$ are the frequencies computed from the first and second 500 turns, respectively. The results across all three models show similar frequency maps. A seventh order resonance $4\nu_y = 7$ can be observed.

To enhance the $z$ variation, one effective approach is to amplify the beam intensity. The beam-beam parameters can be preserved by reducing the $\beta$ functions and maintaining the slice position at $z^* = -\beta/2$. Suppose the beam intensity is increased by a factor of $A$, and concurrently, the $\beta$ functions are reduced by the same factor. This adjustment

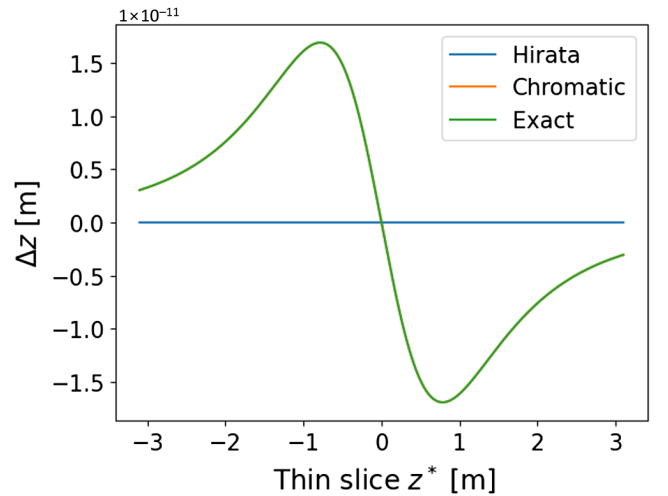

FIG. 2. Longitudinal coordinate variation after projecting back to IP. The legend describes the virtual drift models employed in the simulations: original Hirata's method (blue), chromatic Hamiltonian (orange), and exact Hamiltonian (green). The orange and green curves overlap with each other.

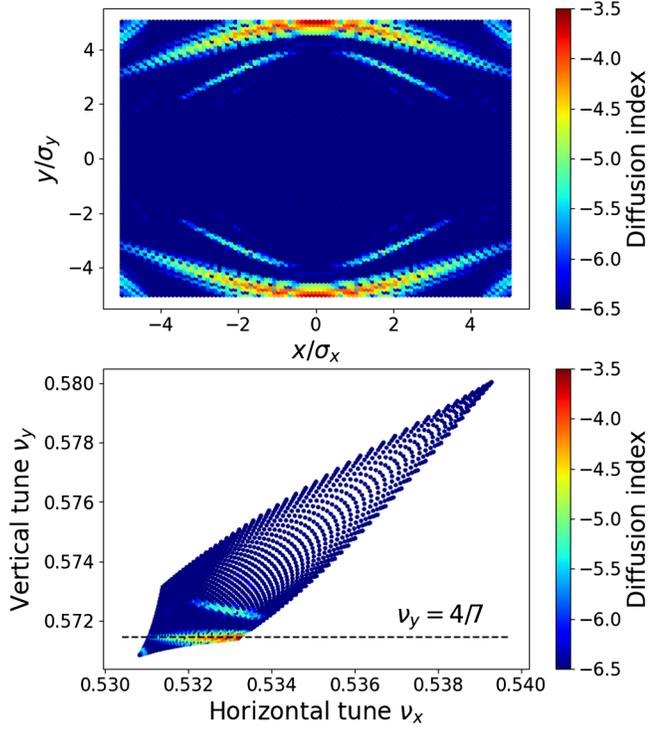

FIG. 3. Frequency map analysis in amplitude space (top) and tune space (bottom).

results in an $A$-fold increase in both beam divergence and the beam-beam kick. According to Eq. (11):

$$\Delta z \approx \frac{S}{2}(2x'\Delta x' + \Delta x'^2 + 2y'\Delta y' + \Delta y'^2), \quad (38)$$

where $x'$, $y'$, $\Delta x'$, and $\Delta y'$ also increase by a factor of $A$, while the distance $S$ between the CP and IP is reduced by a factor of $A$. Consequently, it leads to an $A$-fold increase in $\Delta z$.

In our case, the initial longitudinal coordinates are set to 0. Therefore, the longitudinal action is derived from the energy kick due to beam-beam interactions. When the beam intensity is scaled up, the associated longitudinal energy kick also increases proportionally. Consequently, the change in the longitudinal coordinate, $\Delta z$, remains negligible compared to the longitudinal action. As the $\beta$ functions decrease during this scaling process, the hourglass effect becomes relevant when $\Delta z$ approaches the magnitude of the $\beta$ function. This variance in $\Delta z$ across different models will influence long-term particle diffusion.

Figure 4 displays the frequency maps obtained from different models when the beam intensity is increased by a factor of 100. The reduction in $\beta$ functions leads to fewer particles, indicated in red color, being affected by the seventh order resonance. In the chromatic and exact virtual models, particles close to the center exhibit a higher diffusion index. Therefore, in long-term tracking, the $z$ variance effect should be included.

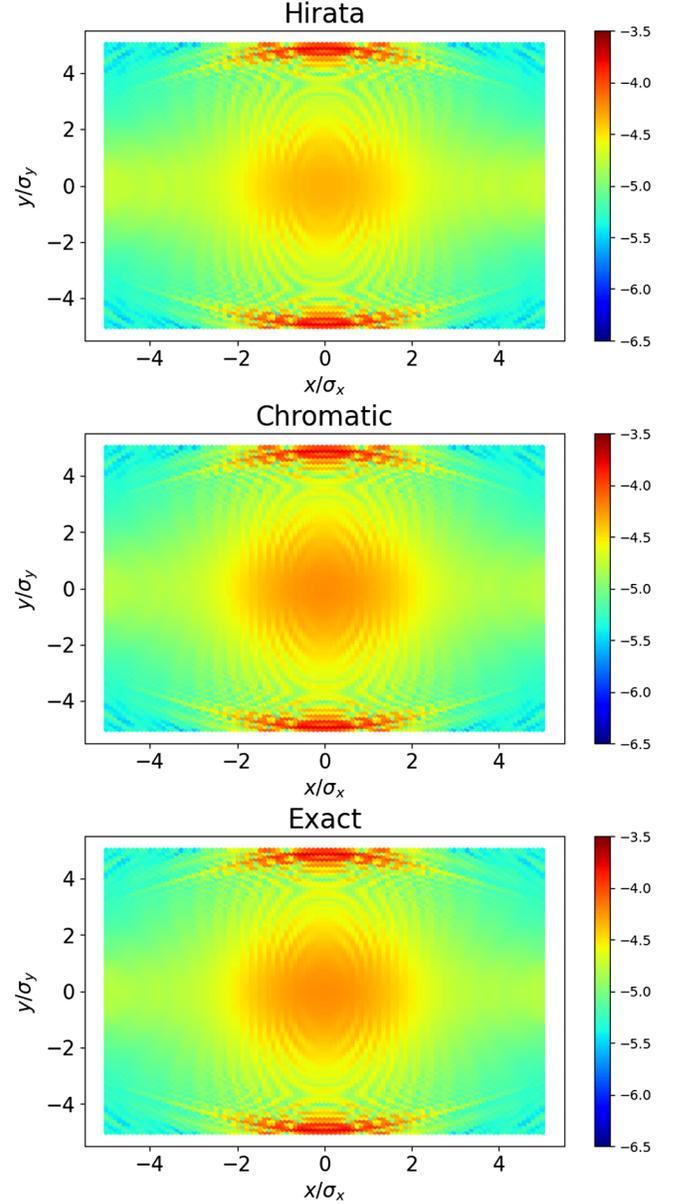

FIG. 4. Frequency map analysis in amplitude space when the beam intensity is scaled by 100 times.

### B. Realistic case: Simulation for EIC

The parameters used in simulations are shown in Table II. The crossing angle is as large as 25 mrad, and the Lorentz transformation is applied to accurately address the crossing angle within the simulation [26]. In all subsequent simulations, the weak proton beam is represented by one million macroparticles. A second order harmonic crab cavity is used to flatten the proton bunch [27]. The one-turn map at IP is represented by the linear betatron map. There is no momentum dispersion and crab dispersion at IP [28]. The electron beam, functioning as the strong beam, exhibits a rigid bi-Gaussian distribution. Parameters for the electron beam, as listed in Table II, are anticipated to be achieved upon reaching equilibrium, accounting for the effects of beam-beam interactions.

TABLE II. Simulation parameters from EIC-CDR [7]. "H" stands for horizontal and "V" denotes vertical below. The electron beam, acting as the strong beam, is characterized by design parameters anticipated to be realized upon achieving equilibrium.

| Parameter | Proton | Electron |
|---|---|---|
| Circumference (m) | 3834 | |
| Energy (GeV) | 275 | 10 |
| Particles per bunch ($10^{11}$) | 0.688 | 1.72 |
| Crossing angle (mrad) | 25.0 | |
| Crab cavity frequency (MHz) | 200.0 | 400.0 |
| $\beta_x^*/\beta_y^*$ (cm) | 80.0/7.20 | 45.0/5.60 |
| rms emittance (H/V) (nm rad) | 11.3/1.00 | 20.0/1.29 |
| rms bunch size (H/V) (μm) | 95.0/8.5 | |
| rms bunch length (cm) | 6.0 | 0.7 |
| rms energy spread ($10^{-4}$) | 6.6 | 5.5 |
| Transverse fractional tune (H/V) | 0.228/0.210 | 0.08/0.14 |
| Synchrotron tune | −0.010 | −0.069 |
| Transverse damping time (turns) | ∞ | 4000 |
| Longitudinal damping time (turns) | ∞ | 2000 |

Figure 5 illustrates the evolution of proton emittance through three distinct mapping approaches. The blue curve represents the original Hirata's mapping, detailed in Eq. (8). Meanwhile, the orange curve displays the results obtained from a modified virtual drift model incorporating the chromatic Hamiltonian, as derived from Eqs. (22) and (23). Finally, the green curve showcases the outcomes derived from employing a modified virtual drift model with the exact Hamiltonian, as outlined in Eqs. (35) and (36).

The emittance growth rate is determined through a linear fit of the final 50% of the tracking data. To more clearly illustrate the trend of emittance evolution, the data are averaged over every 1000 turns. In the context of the EIC design, vertical emittance growth is of particular concern. For the baseline design, it is imperative that the vertical emittance growth does not surpass 20%/h, a limit set by the Strong Hadron Cooling requirements. As depicted in Fig. 5, the vertical emittance growth observed across the three examined approaches shows negligible differences. This outcome is expected, given that the z variation effect constitutes a higher-order effect, and the parameters listed in Table II have been fine-tuned to minimize vertical emittance growth.

Figure 6 presents the normalized relative error derived from tracking data, offering valuable insights into our enhanced model. The linear progression of relative error over time suggests that the proton beam is free from significant resonance phenomena. Furthermore, the error's magnitude is notably small, on the order of $10^{-5}$, affirming the effectiveness of Hirata's original mapping across a majority of scenarios. Additionally, a comparison between the modified drift model, which integrates the chromatic Hamiltonian, and the exact Hamiltonian model reveals a remarkable degree of consistency, highlighting the robustness of our approach.

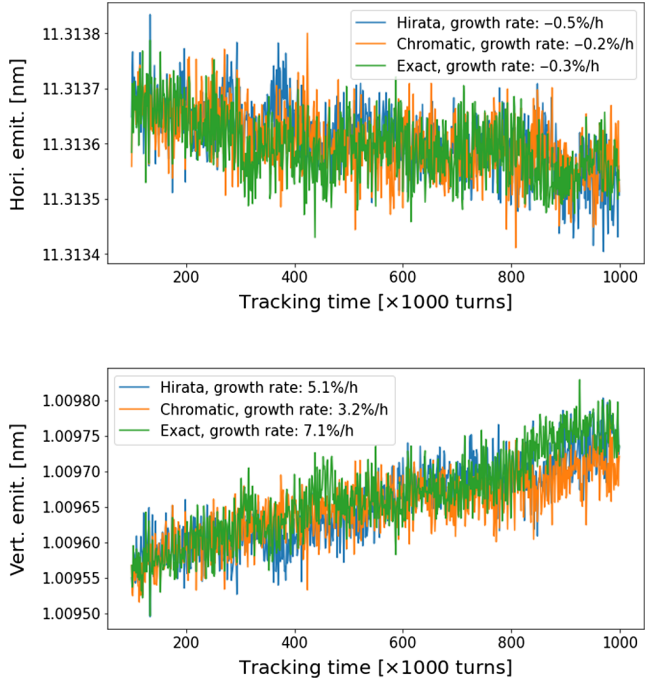

FIG. 5. Evolution of proton emittance depicted through weak-strong simulations: top (horizontal plane) and bottom (vertical plane). The legend describes the virtual drift models employed in the simulations: original Hirata's method (blue), chromatic Hamiltonian (orange), and exact Hamiltonian (green).

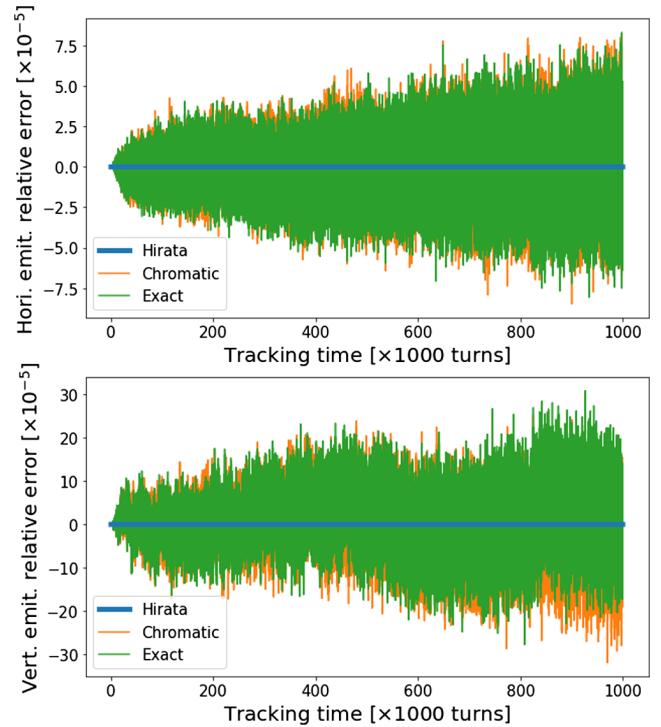

FIG. 6. Relative error of proton emittance in weak-strong simulations: top (horizontal plane), bottom (vertical plane). The legend describes the virtual drift models employed in the simulations: original Hirata's method (blue), chromatic Hamiltonian (orange), and exact Hamiltonian (green).

It is important to note that a relative error on the order of $10^{-5}$ or $10^{-4}$ per turn may not be considered negligible when compared to the magnitude of random diffusion processes, such as intrabeam scattering (IBS). In the context of the EIC, where the horizontal or longitudinal IBS growth time is approximately 2–3 h, the relative amplitude of IBS diffusion is about $10^{-7}$ per turn. From this perspective, the inclusion of the $z$ variation effect in beam-beam interaction becomes essential in the design of the hadron ring, when addressing the interplay between beam-beam interaction and the IBS.

In accelerator physics, it is a well-established fact that minor deviations can be exponentially magnified over time for chaotic particle motion. This scenario typically arises when there is an overlap of multiple resonance lines within the tune space. Specifically, for the EIC beam-beam simulation, high-order synchro-betatron resonances are present within the footprint, as detailed in [29]. Figure 7 illustrates the tracking outcomes when the strong electron beam is artificially displaced by one $\sigma_x$ in the horizontal plane. This displacement excites the higher-order synchro-betatron resonances, leading to a noticeable increase in vertical emittance across three distinct simulation models.

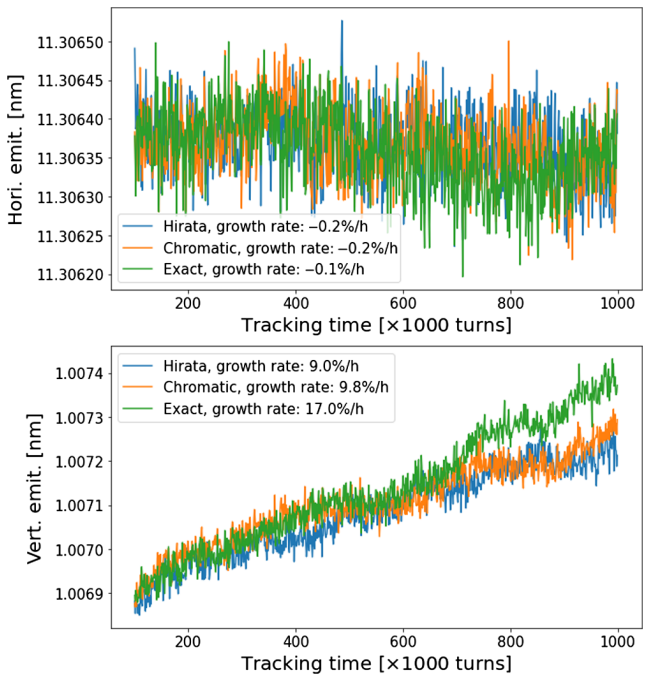

FIG. 7. Proton emittance evolution in weak-strong simulation with strong electron beam offset by one $\sigma_x$ in the horizontal plane: top (horizontal plane) and bottom (vertical plane). Here, $\sigma_x$ represents the rms horizontal size of the electron beam at the IP. The legend follows the same conventions as in previous figures.

Incorporating a realistic lattice model inevitably introduces more higher-order resonances, particularly at larger amplitudes. Figure 7 underscores the necessity of incorporating $z$ variation in the dynamic aperture study.

## V. CONCLUSION AND OUTLOOK

In weak-strong simulations for beam-beam interactions, Hirata's model is widely used, which offers an approximate, yet symplectic mapping that accounts for energy changes and bunch length effects. However, it operates under the assumption of constant longitudinal coordinates ($z$), thereby excluding $z$ variation from its considerations.

In the design of hadron rings, it is essential to track particles across millions of turns, making it impractical to disregard $z$-variation during beam-beam interactions. Building upon Hirata's pioneering work, we propose two new models designed to refine the virtual drift process. It can be shown that these models are symplectic and accurately integrate the $z$-variation effect at the lowest order.

Simulations are performed to benchmark these three distinct models. Although the result indicates that $z$ variation is a higher-order effect that does not substantially alter the emittance growth rate under the specific design considerations of the EIC, the discrepancy among the models is significant when contrasted with the amplitude of IBS diffusion. This distinction becomes even more critical when considering realistic lattice models that involve higher-order resonances, underscoring the importance of accounting for $z$ variation.

Moreover, the approach discussed in Sec. III B can be readily adapted to include external fields, such as those from a detector solenoid, without having to assume that the $z$ coordinate remains constant in the presence of such fields.

We have to emphasize that the enhanced models presented in Sec. III, along with Hirata's original model, should be regarded as approximations. These models presuppose that the transformations between the IP and the CP adhere to symplectic principles with respect to accelerator coordinates, a notion that deviates from physical reality as discussed in Sec. II B. Nonetheless, we also recognize that in the absence of dissipation and diffusion, real-world physical processes must inherently conform to symplectic characteristics.

An alternate approach involves standard tracking (no energy kick) between IP and CP with the distance between IP and CP determined by the longitudinal coordinates $z$ of the test particle and the $z^*$ of the opposing slice along with the trajectory of the test particle. This is coupled with a beam-beam interaction at the CP that includes an energy kick due to the slingshot effect and an energy kick due to the varying shape of the slice. This strategy has been applied within the framework of Bmad [24].

Shatilov and Zobov in [30] also employed this strategy: they used the real drift map and modify the beam-beam kick accordingly. In their paper, the additional energy kick is attributed to $v_x B_y - v_y B_x$, where $v_{x,y}$ are transverse velocities and $B_{x,y}$ are transverse magnetic field from beam-beam interaction. In Bmad and in Appendix C, this energy kick is derived by analogy to the slingshot effect. Both treatments yield consistent formulas under the high-relativistic approximation.

The advantage of this method is that it more closely mirrors what is actually happening. The disadvantage is that symplecticity of the overall map is not assured. Future work will entail a detailed comparison between the methodologies employed by Hirata and the one used by Bmad and Shatilov.

## ACKNOWLEDGMENTS

The authors would like to thank J. Scott Berg, M. Blaskiewicz, G. Hoffstaetter, and J. Qiang for insightful discussions. This work was supported by Brookhaven Science Associates, LLC under Contract No. DE-SC0012704 with the U.S. Department of Energy. This manuscript has been authored in part by UT-Battelle, LLC, under Contract No. DE-AC05-00OR22725 with the U.S. Department of Energy. Y. H. is supported by the adjoint appointment agreement between Michigan State University and Brookhaven National Laboratory.

## APPENDIX A: HAMILTONIAN AND CANONICAL VARIABLES

In accordance with Forest [31], the Hamiltonian for a relativistic particle in an external magnetic vector potential **A**, navigating through a curved coordinate system of radius $\rho$, is formulated as

$$K = -\left(1 + \frac{x}{\rho}\right)\sqrt{(1+\delta)^2 - p_x^2 - p_y^2} + \frac{x}{\rho} + \frac{x^2}{2\rho^2} - \frac{A_s(x,y)}{B_0 \rho}. \quad (A1)$$

Here, $\delta = (P - P_0)/P_0$ represents the relative momentum deviation from the reference momentum $P_0$, and $p_x$, $p_y$ denote canonical momenta normalized by $P_0$.

The canonical variables in this context are represented by two equivalent sets:

$$(x, p_x, y, p_y, -l, \delta) \quad \text{or} \quad (x, p_x, y, p_y, \delta, l), \quad (A2)$$

with $l$ denoting the particle's path length. The reference particle's design path length, $s$, serves as the independent variable in the Hamiltonian $K$ expressed in Eq. (A1).

The generating function

$$F_3(z, \delta; s) = (s - z)\delta \quad (A3)$$

yields the transformations:

$$-l = -\frac{\partial F_3}{\partial \delta} = z - s \Rightarrow z = s - l \quad (A4)$$

and

$$p_z = -\frac{\partial F_3}{\partial z} = \delta \quad (A5)$$

leading to a new Hamiltonian formulation:

$$\begin{aligned} H &= 1 + K + \frac{\partial F_3}{\partial s} \\ &= 1 + p_z - \left(1 + \frac{x}{\rho}\right)\sqrt{(1+p_z)^2 - p_x^2 - p_y^2} \\ &\quad + \frac{x}{\rho} + \frac{x^2}{2\rho^2} - \frac{A_s(x,y)}{B_0 \rho}. \end{aligned} \quad (A6)$$

The corresponding canonical variables are

$$\left(x, p_x = \frac{P_x}{P_0}, y, p_y = \frac{P_y}{P_0}, z = s - l, p_z = \frac{P - P_0}{P_0}\right). \quad (A7)$$

Similar to [26], a constant 1 is added to the Hamiltonian. It has no effect on beam dynamics.

For a particle in a drift space, where $\rho = \infty, A_s = 0$, the Hamiltonian simplifies to

$$H_{\text{drift}} = 1 + p_z - \sqrt{(1+p_z)^2 - p_x^2 - p_y^2}. \quad (A8)$$

## APPENDIX B: BEAM-BEAM KICK FOR BI-GAUSSIAN DISTRIBUTION

For a bi-Gaussian particle distribution, the beam-beam potential $U$ is given by

$$U(x, y; \sigma_x, \sigma_y) = \frac{Q_1 Q_2 N r_0}{\gamma_0} \int_0^\infty du \frac{\exp\left(-\frac{x^2}{2\sigma_x^2 + u} - \frac{y^2}{2\sigma_y^2 + u}\right)}{\sqrt{2\sigma_x^2 + u}\sqrt{2\sigma_y^2 + u}}, \quad (B1)$$

where $N$ is the total particle number, $r_0 = e^2/(4\pi\epsilon_0 m c^2)$ the classical radius, $\gamma_0$ the relativistic factor of the test particle, $Q_{1,2}$ the charge numbers of particles from two colliding bunches, and $\sigma_{x,y}$ are the rms beam sizes of the strong slice at the CP. The parameters $\sigma_x$ and $\sigma_y$ are not constants but vary as functions of the distance $S$ from the IP to the CP.

The first order differential was derived by Bassetti and Erskine [10]

$$U_y + iU_x = -\frac{Q_1 Q_2 N r_0}{\gamma_0}\sqrt{\frac{2\pi}{\sigma_x^2 - \sigma_y^2}}\left[w(z_2) - w(z_1)\exp\left(-\frac{x^2}{2\sigma_x^2} - \frac{y^2}{2\sigma_y^2}\right)\right], \quad (B2)$$

where

$$U_x \equiv \frac{\partial U(x,y)}{\partial x}, \qquad U_y \equiv \frac{\partial U(x,y)}{\partial y}$$

$$z_1 = \frac{\frac{\sigma_y}{\sigma_x}x + i\frac{\sigma_x}{\sigma_y}y}{\sqrt{2(\sigma_x^2 - \sigma_y^2)}}, \qquad z_2 = \frac{x + iy}{\sqrt{2(\sigma_x^2 - \sigma_y^2)}}.$$

In Eq. (B2), $w(z)$ is the Faddeeva function defined as

$$w(z) \equiv \exp(-z^2)\left(1 + \frac{2i}{\sqrt{\pi}}\int_0^z dt\, e^{t^2}\right). \tag{B3}$$

Its derivative is given by

$$w'(z) = \frac{2i}{\sqrt{\pi}} - 2zw(z). \tag{B4}$$

The parameters $\sigma_{x,y}$ vary as functions of the longitudinal coordinate $z$. The derivative $U_z$ was first given by Hirata in [23]:

$$\begin{aligned}U_z &= \frac{\partial U}{\partial \sigma_x}\frac{\partial \sigma_x}{\partial z} + \frac{\partial U}{\partial \sigma_y}\frac{\partial \sigma_y}{\partial z}\\ &= \sigma_x U_{xx}\frac{\partial \sigma_x}{\partial z} + \sigma_y U_{yy}\frac{\partial \sigma_y}{\partial z}.\end{aligned} \tag{B5}$$

Taking the partial derivatives on both sides of Eq. (B2), the second order derivatives can be obtained

$$U_{xx} = -\frac{xU_x + yU_y}{\sigma_x^2 - \sigma_y^2} - \frac{2Q_1Q_2Nr_0}{\gamma_0(\sigma_x^2 - \sigma_y^2)}$$
$$\times\left[1 - \frac{\sigma_y}{\sigma_x}\exp\left(-\frac{x^2}{2\sigma_x^2} - \frac{y^2}{2\sigma_y^2}\right)\right], \tag{B6}$$

$$U_{yy} = \frac{xU_x + yU_y}{\sigma_x^2 - \sigma_y^2} + \frac{2Q_1Q_2Nr_0}{\gamma_0(\sigma_x^2 - \sigma_y^2)}$$
$$\times\left[1 - \frac{\sigma_x}{\sigma_y}\exp\left(-\frac{x^2}{2\sigma_x^2} - \frac{y^2}{2\sigma_y^2}\right)\right]. \tag{B7}$$

As a result, the momentum changes in Eq. (6) for a strong slice with bi-Gaussian distribution are fully resolved.

## APPENDIX C: ENERGY CHANGE DUE TO A MOVING MAGNET

The four-momentum of a test particle in the laboratory frame is denoted as $(E/c, \mathbf{P})$, where $E$ represents the energy of the particle, $c$ is the speed of light, and $\mathbf{P}$ is its momentum vector. When transitioning to the rest frame of the moving magnet, the energy of the test particle undergoes a transformation as per the principles of Lorentz transformation. This transformation is articulated as

$$\frac{\bar{E}}{c} = \gamma^*\left(\frac{E}{c} + \beta^* P \cos\theta\right), \tag{C1}$$

where $\beta^* = v^*/c$ is the magnet's velocity normalized by the speed of the light, and $\gamma^*$ is the corresponding Lorentz factor. $\theta$ represents the angle between the particle's momentum vector and the opposite direction of the magnet's motion. The notation $\bar{E}$ indicates the energy measured in the magnet's rest frame.

In this rest frame, the test particle's trajectory is altered while its energy is conserved. Therefore,

$$\gamma^*\left(\frac{E_2}{c} + \beta^* P_2 \cos\theta_2\right) = \gamma^*\left(\frac{E_1}{c} + \beta^* P_1 \cos\theta_1\right). \tag{C2}$$

For relativistic scenarios, we approximate

$$P_2 \approx \frac{E_2}{c}, \qquad P_1 \approx \frac{E_1}{c}, \tag{C3}$$

where subscripts "1" and "2" denote the states before and after the interaction with the magnet, respectively.

Substituting these approximations into Eq. (C2) yields the energy change:

$$\Delta E = E_2 - E_1 = \frac{E_1\beta^*(\cos\theta_1 - \cos\theta_2)}{1 + \beta^*\cos\theta_2}. \tag{C4}$$

Utilizing the paraxial approximation where $\theta_{1,2} \approx 0$ and considering the high relativistic limit where $\beta^* \approx 1$, the formula simplifies further to

$$\Delta E \approx \frac{E_1}{4}(\theta_2^2 - \theta_1^2). \tag{C5}$$

---

[1] M. Benedikt and F. Zimmermann, Proton colliders at the energy frontier, Nucl. Instrum. Methods Phys. Res., Sect. A **907**, 200 (2018).

[2] G. Heinrich, Collider physics at the precision frontier, Phys. Rep. **922**, 1 (2021).

[3] A. Accardi, J. Albacete, M. Anselmino, N. Armesto, E. Aschenauer, A. Bacchetta, D. Boer, W. Brooks, T. Burton, N. B. Chang *et al.*, Electron-ion collider: The next QCD frontier: Understanding the glue that binds us all, Eur. Phys. J. A **52**, 1 (2016).

[4] K. Ohmi, M. Tawada, Y. Cai, S. Kamada, K. Oide, and J. Qiang, Beam-beam limit in $e^+e^-$ circular colliders, Phys. Rev. Lett. **92**, 214801 (2004).

[5] J. Beebe-Wang and S. Zhang, Observation and simulation of beam-beam induced emittance growth in RHIC, Brookhaven National Laboratory, Upton, NY, Technical Report No. BNL-81793-2009-CP, 2009.

[6] S. Paret and J. Qiang, Simulation of beam-beam induced emittance growth in the HL-LHC with crab cavities, arXiv:1410.5964.

[7] F. Willeke and J. Beebe-Wang, Electron ion collider conceptual design report 2021, edited by Thomas Jefferson *et al.*, Brookhaven National Laboratory, Upton, NY, Technical Report No. BNL-221006-2021-FORE; TRN: US2215154, 2021, https://www.osti.gov/biblio/1765663.

[8] F. Willeke, The HERA lepton–proton collider, in *Challenges and Goals for Accelerators in the XXI Century* (World Scientific Publishing Company, Singapore, 2016), Chap. 15, pp. 225–242.

[9] S. Peggs and R. Talman, Beam-beam luminosity limitation in electron-positron colliding rings, Phys. Rev. D **24**, 2379 (1981).

[10] M. Bassetti and G. A. Erskine, Closed expression for the electrical field of a two-dimensional Gaussian charge, Technical Report No. CERN-ISR-TH-80-06; ISR-TH-80-06, 1980, https://cds.cern.ch/record/122227?ln=en.

[11] K. Ohmi, Simulation of beam-beam effects in a circular $e^+e^-$ collider, Phys. Rev. E **62**, 7287 (2000).

[12] J. Qiang, M. A. Furman, and R. D. Ryne, A parallel particle-in-cell model for beam–beam interaction in high energy ring colliders, J. Comput. Phys. **198**, 278 (2004).

[13] Y. Zhang, K. Ohmi, and L. Chen, Simulation study of beam-beam effects, Phys. Rev. ST Accel. Beams **8**, 074402 (2005).

[14] Y. Luo, Y. Hao, J. Qiang, Y. Roblin, F. Willeke, H. Zhang *et al.*, Simulation challenges for eRHIC beam-beam study, in *Proceedings of 10th International Particle Accelerator Conference, IPAC-2019, Melbourne, Australia* (JACoW, Geneva, Switzerland, 2019), pp. 785–787.

[15] D. Xu, Y. Luo, C. Montag, Y. Hao, and J. Qiang, Numerical noise study in EIC beam-beam simulations, in *Proceedings of the 12th International Particle Accelerator Conference, IPAC-2021, Campinas, SP, Brazil* (JACoW, Geneva, Switzerland, 2021), Vol. 2021.

[16] D. Xu, Y. Hao, Y. Luo, C. Montag, and J. Qiang, Model parameters determination in EIC strong-strong simulation, in *Proceedings of the 5th North American Particle Accelerator Conference, NAPAC-2022, Albuquerque, NM* (JACoW, Geneva, Switzerland, 2022), pp. 9–11.

[17] Y. Luo, J. Berg, M. Blaskiewicz, W. Fischer, B. Gamage, X. Gu, G. Hoffstaetter, H. Huang, J. Kewisch, H. L. III, C. Montag, V. Morozov, E. Nissen, S. Peggs, V. Ptitsyn, J. Qiang, T. Satogata, F. Willeke, and D. Xu, Summary of numerical noise studies for electron-ion collider strong-strong beam-beam simulation, in *Proceedings of the 13th International Particle Accelerator Conference, IPAC-2022, Bangkok, Thailand* (JACoW, Geneva, Switzerland, 2022), pp. 1931–1934.

[18] Y. Luo, J. Berg, M. Blaskiewicz, W. Fischer, X. Gu, Y. Hao, H. Huang, H. L. III, C. Montag, V. Morozov, E. Nissen, R. Palmer, S. Peggs, V. Ptitsyn, J. Qiang, T. Satogata, F. Willeke, and D. Xu, Numerical noise error of particle-in-cell poisson solver for a flat Gaussian bunch, in *Proceedings of the 13th International Particle Accelerator Conference, IPAC-2022, Bangkok, Thailand* (JACoW, Geneva, Switzerland, 2022), pp. 1939–1941.

[19] Y. Luo, J. Berg, W. Fischer, X. Gu, Y. Hao, H. L. III, C. Montag, V. Morozov, S. Peggs, V. Ptitsyn, J. Qiang, T. Satogata, H. Witte, and D. Xu, Dynamic aperture evaluation for EIC Hadron storage ring with crab cavities and IR nonlinear magnetic field errors, in *Proceedings of the 13th International Particle Accelerator Conference, IPAC-2022, Bangkok, Thailand* (JACoW, Geneva, Switzerland, 2022), pp. 1927–1930.

[20] D. Xu, M. Blaskiewicz, Y. Luo, D. Marx, C. Montag, and B. Podobedov, Effect of electron orbit ripple on proton emittance growth in EIC, in *Proceedings of the 14th International Particle Accelerator Conference, IPAC-2023, Venice, Italy* (JACoW, Geneva, Switzerland, 2023), pp. 108–111.

[21] H. Huang, Y. Zhang, T. Satogata, V. Morozov, Y. L. F. Lin, D. Xu, and Y. Hao, Quantifying effects of crab cavity rf phase noise on transverse emittance in EIC Hadron storage ring, in *Proceedings of the 14th International Particle Accelerator Conference, IPAC-2023, Venice, Italy* (JACoW Publishing, Geneva, Switzerland, 2023), pp. 2399–2401.

[22] Y. Hao, Y. Luo, D. Xu, V. Morozov, and H. Huang, Validation and countermeasures of vertical emittance growth due to crab cavity noise in a horizontal crab-crossing scheme, in *Proceedings of the 14th International Particle Accelerator Conference, IPAC-2023, Venice, Italy* (JACoW, Geneva, Switzerland, 2023), pp. 97–100.

[23] K. Hirata, H. W. Moshammer, and F. Ruggiero, A symplectic beam-beam interaction with energy change, Part. Accel. **40**, 205 (1992), https://cds.cern.ch/record/243013/files/p205.pdf.

[24] D. Sagan, *The Bmad Reference Manual* (2024), https://www.classe.cornell.edu/bmad/manual.html.

[25] K. Akai, K. Furukawa, and H. Koiso, Superkekb collider, Nucl. Instrum. Methods Phys. Res., Sect. A **907**, 188 (2018).

[26] K. Hirata, Analysis of beam-beam interactions with a large crossing angle, Phys. Rev. Lett. **74**, 2228 (1995).

[27] D. Xu, Y. Hao, Y. Luo, C. Montag, and J. Qiang, Study of harmonic crab cavity in EIC beam-beam simulations, in *Proceedings of the 12th International Particle Accelerator Conference, IPAC-2021, Campinas, SP, Brazil* (JACoW, Geneva, Switzerland, 2021), pp. 2595–2597.

[28] D. Xu, Y. Luo, and Y. Hao, Combined effects of crab dispersion and momentum dispersion in colliders with local crab crossing scheme, Phys. Rev. Accel. Beams **25**, 071002 (2022).

[29] D. Xu, Y. Hao, Y. Luo, and J. Qiang, Synchro-betatron resonance of crab crossing scheme with large crossing angle and finite bunch length, Phys. Rev. Accel. Beams **24**, 041002 (2021).

[30] D. Shatilov and M. Zobov, Beam-beam collisions with an arbitrary crossing angle: Analytical tune shifts, tracking algorithm without Lorentz boost, crab-crossing, in *ICFA Beam Dynamics Newsletter* (2005), pp. 99–109, https://icfa-usa.jlab.org/archive/newsletter/icfa_bd_nl_37.pdf.

[31] E. Forest, Geometric integration for particle accelerators, J. Phys. A **39**, 5321 (2006).